\documentclass[review,12pt]{elsarticle}

\journal{Computer Physics Communications}

\usepackage[british]{babel}
\usepackage{hyphenat}

\usepackage{lineno,hyperref}
\modulolinenumbers[5]

\usepackage{graphicx,amssymb}
\usepackage{color}

\begin{document}

\begin{frontmatter}

\title{Validation of moment tensor potentials for fcc and bcc metals using EXAFS spectra}


\author[SKL]{Alexander\ V.\ Shapeev\corref{cor1}}
\ead{alexander@shapeev.com}
\cortext[cor1]{Corresponding author}

\author[ISSP]{Dmitry\ Bocharov}
\ead{Dmitrijs.Bocarovs@cfi.lu.lv}

\author[ISSP]{Alexei\ Kuzmin\corref{cor2}}
\ead{a.kuzmin@cfi.lu.lv}
\cortext[cor2]{Corresponding author}

\address[SKL]{Skolkovo Institute of Science and Technology, Skolkovo Innovation Center, 3, Moscow 143026, Russian Federation}
 
\address[ISSP]{Institute of Solid State Physics, University of Latvia,
Kengaraga Street 8, LV-1063 Riga, Latvia}


\begin{abstract}
Machine-learning potentials for materials, namely the moment tensor potentials (MTPs), were validated using experimental EXAFS spectra for the first time. The MTPs for four metals (bcc W and Mo, fcc Cu and Ni) were obtained by the active learning algorithm of fitting to the results of the calculations using density functional theory (DFT).  The MTP accuracy was assessed by comparing  metal K-edge EXAFS spectra obtained experimentally and computed from the results of molecular dynamics (MD) simulations. The sensitivity of the method to various aspects of the MD and DFT models was demonstrated using Ni as an example. Good agreement was found for W, Mo and Cu using the recommended PAW pseudopotentials, whereas a more accurate pseudopotential with 18 valence electrons was required for Ni to achieve a similar agreement. The use of EXAFS spectra allows one to estimate the MTP ability in reproducing  both average and dynamic atomic structures. 
\end{abstract}

\end{frontmatter}


 \newpage


Machine-learning potentials are becoming an increasingly useful tool for atomistic modeling, combining the best features of the more traditional methods, empirical potentials and first-principles (or quantum-mechanical) models.
The favorable features of the latter include the fact that they do not rely on empirical information about the material they are employed to model (hence their name), universality in the sense of the ability to simultaneously describe a wide class of different compounds, quantitative accuracy sufficient to make predictions for newly designed materials, and acceptable computational efficiency at least when calculating the zero-temperature properties of materials.
The advantage of empirical potentials, on the other hand, is that they are many orders of magnitude more efficient than the first-principles models because they do not need to resolve the electronic structure for every given atomistic structure which may be important when simulating systems over very large space and time scales.
Empirical potentials postulate a certain functional form of the energy of interatomic interaction with one or more free parameters that are optimized by requiring that the potential reproduces certain known (usually quantum-mechanical) data.
Due to the simple functional form, the accuracy of empirical potentials in many cases is only sufficient to reveal atomistic mechanisms leading to a certain experimentally observed property, but not making a de novo computational prediction.
That is why the free parameters of the potentials are fitted to available experimental data in addition to (or in the past---instead of) quantum-mechanical data.

The quality of quantum-mechanical calculations is usually assessed by directly comparing them to the experimental data.
There is extensive data in the literature of such comparison for zero-temperature properties such as lattice constants or elastic moduli or phonon properties  \cite{grabowski2007-ab-initio,Hao2012-lattice-constant-with-ZPE}.
Fewer works are comparing quantum-mechanical calculations of finite-temperature properties with the experimental values \cite{zhu2020-melting-al-ni,Haskins2017}.

Machine-learning potentials have been extensively used to calculate high-temperature materials properties \cite{novikov2020-mlip,smith2021-aluminum,pun2019-pinn,park2020-gnn,shapeev2020-elinvar,jinnouchi2019-kresse-on-the-fly,rowe2020-gap-carbon,cheng2019-water}.
When such property has been computed with DFT with the same convergence parameters (such as the $k$-point mesh or the plane-wave energy cutoff), the difference between the results computed by DFT and machine-learning potentials differ much less than the DFT error --- e.g., the melting point of Al computed with the same DFT parameters was different by only 2 K \cite{zhu2020-melting-al-ni,novikov2020-mlip}, whereas both were about 40 K smaller than the experimental value.
Such a property of machine-learning potentials make them a good tool for the experimental validation of DFT functionals: indeed, comparing the results of a well-trained machine-learning potential to the experimental data is essentially the same as comparing the underlying DFT functional to these experimental data.

Among different experimental techniques, the extended X-ray absorption fine structure (EXAFS) spectroscopy is well suited for validation of the results of molecular dynamics (MD) simulations \cite{Kuzmin2020}, since the EXAFS spectrum offers a much richer set of information about atomistic dynamics of material than a single quantity like melting point.
Indeed, EXAFS spectra contain information on the local structure and vibrational dynamics around an absorbing element so that crystalline, nanocrystalline or disordered materials can be studied equally well over a wide range of in situ and operando conditions \cite{Bordiga2013,Kuzmin2014exafs,Mino2018,Grunert2020,Timoshenko2021review}.
Moreover,  EXAFS spectra depend on pair and many-atom distribution functions, thus providing not only radial but also angular sensitivity \cite{Filipponi1995a,Rehr2000,Natoli2003}. Note that pair distribution functions (PDFs) contribute  to the EXAFS spectrum via the single-scattering (SS) processes, whereas many-atom distribution functions give rise to multiple-scattering (MS) processes \cite{Rehr2000,Natoli2003}.

The characteristic time of the photoabsorption event (10$^{-15}$--10$^{-16}$~s) is shorter than that of thermal vibrations (10$^{-13}$--10$^{-14}$~s), therefore, the atoms can be regarded as frozen at their instantaneous positions, and the measured EXAFS spectrum is the average over all atomic configurations which appear during the time of the experiment. 
This situation is similar to that occurring during the MD simulation, when a set of instantaneous atomic configurations (`snapshots') is generated,
giving a view of the dynamic evolution of the material structure. EXAFS spectrum 
can be calculated for each snapshot and, finally, the configuration-averaged EXAFS spectrum 
can be compared with the experimental one. Performing MD simulations using different models of interatomic potentials, several theoretical EXAFS spectra can be obtained, and the agreement between each of them and the experimental EXAFS spectrum can be used to evaluate the model accuracy. The MD-EXAFS method was successfully used in the past for validation of the empirical interatomic potentials in different oxides \cite{Anspoks2011,Timoshenko2014,Kuzmin2016,Jonane2016a,Bocharov2017} and ab initio molecular dynamics simulations of ScF$_3$ \cite{Bocharov2020scf3}.
 
In this study, we evaluate the quality of machine learning potentials reproducing the EXAFS spectra for two bcc (W and Mo) and two fcc (Cu and Ni) metals.
The experimental data of metal foils were taken from our previous studies \cite{Jonane2020,Bakradze2020}. Among these four elements, Ni is an ideal choice for testing how different DFT settings (pseudopotentials, convergence parameters, spin-polarized calculations) affect the results of ab initio-trained machine-learning potentials.


The NVT MD simulations were performed using Moment Tensor Potentials (MTPs) \cite{shapeev2016moment,gubaev2019-alloys} belonging to a class of machine-learning interatomic potentials with favorable accuracy-vs-efficiency balance \cite{zuo2020-benchmark}.
The MTPs were fitted to density functional theory (DFT) data on-the-fly using an active learning algorithm as implemented in the MLIP-2 software package \cite{novikov2020-mlip}.
Sets of atomic configurations obtained from the MD simulations were used to calculate
the configuration-averaged Ni, Cu, and Mo K-edge and W L$_3$-edge EXAFS spectra $\chi(k)$ ($k$ is the photoelectron wavenumber) using the MD-EXAFS method \cite{Kuzmin2014,Kuzmin2020}.
A detailed description of all steps of our simulations can be found in the Supplementary Material. 


In Fig.\ \ref{fig:1} we show the comparison between the experimental and calculated W L$_3$-edge and Mo, Cu and Ni  K-edges EXAFS $\chi(k)k^2$ spectra, their Fourier transforms (FTs) and the resulting pair distribution functions (PDFs) $g(r)$ for bcc W/Mo and fcc Cu/Ni metals at 300~K. Here the results for Ni correspond to those derived from non-magnetic DFT calculations. The multiple-scattering (MS) contributions to EXAFS spectra and their FTs are shown by blue curves. The large amplitude of the MS signals indicates their importance in both $k$ and $R$ spaces, as one can expect for cubic lattices with many linear atomic chains.  In $k$-space, the MS contributions extend over the entire range, whereas in $R$-space the MS peaks appear at long distances above 3--3.5~\AA\ due to the large lengths of the MS paths.

\begin{figure}[t]
	\begin{center}
		\includegraphics[width=0.95\linewidth]{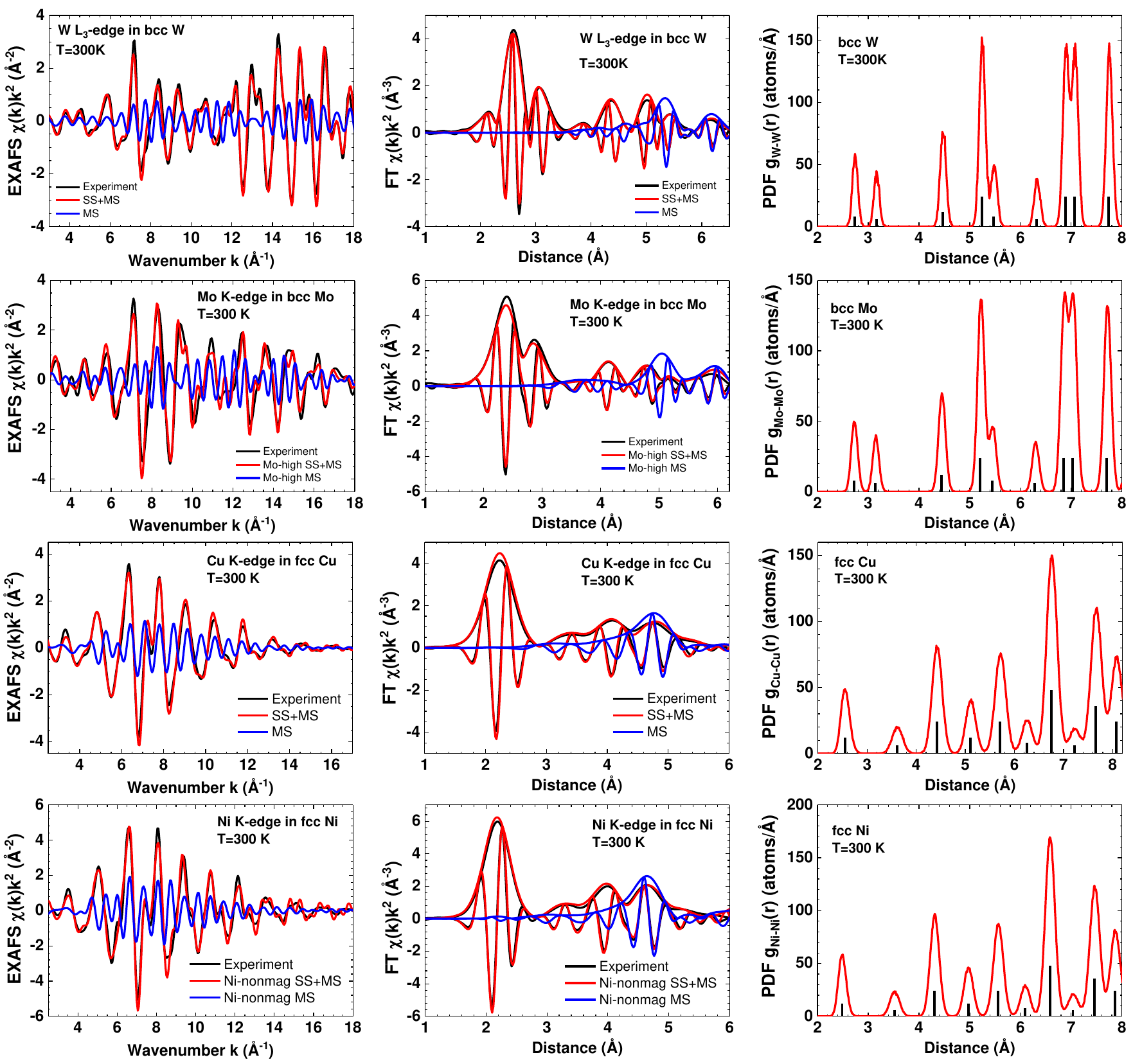}					
	\end{center}
	\caption{Comparison of the results of the MD-EXAFS simulations for bcc W and Mo and fcc Cu and Ni metals. Experimental and calculated EXAFS $\chi(k)k^2$ spectra at 300~K and their Fourier transforms are shown in the left and middle panels, respectively. The total spectra including both SS and MS contributions are shown by red lines, whereas the MS contributions are shown by blue lines. The pair distribution functions (PDFs) $g(R)$ (solid curves) are shown in the right panels together with the positions of crystallographic shells (vertical bars). }
	\label{fig:1}
\end{figure}

EXAFS spectra of the bcc and fcc phases have characteristic shapes in $k$-space and can also be well distinguished in FTs. There is one well separated main peak at 2.2~\AA\ in FTs of the fcc structures of Cu and Ni due to the nearest group of 12 atoms, whereas there are two groups of 8 and 6 atoms contributing to the two peaks in the range of 2--3.5~\AA\ in FTs of the bcc structures of W and Mo.      

The position of MD simulated coordination shells in PDFs $g(R)$  are in good agreement with crystallographic data (see vertical bars in the PDF plots in Fig.\ \ref{fig:1}). Similar to FTs,
the bcc and fcc phases can be easily distinguished from the position of the first two peaks in the PDFs. In the bcc phase, the proximity of the lattice constants of W and Mo results in a small separation ($\sim$0.43~\AA) between the peaks of the first and second coordination shells located at about 2.73~\AA\  and 3.16~\AA. In the fcc phase (for Cu and Ni), the separation between the two shells increases significantly to about 1.0~\AA.       

The overall agreement between the experimental and calculated EXAFS spectra $\chi(k)k^2$  is good in both $k$ and $R$ spaces for all four metals. 
Note that no fitting structural parameters were used in the EXAFS calculations so that the shapes of EXAFS spectra are  unambiguously determined by the results of MTP-based MD simulations.  A more accurate comparison of the obtained results indicates that while MD simulations reproduce  well the average crystallographic lattice, some deviations between EXAFS spectra can be still observed due to inaccuracies in vibrational dynamics. For example, the MTP-predicted dynamics is slightly softer (stiffer) in the first coordination shell of Mo (Cu) that is observed as a difference between the experiment and theory in the first peak amplitude in FTs. Slightly stiffer nearest neighbour interactions are
present also in Ni, however, there is additionally a small divergence in FTs around 4~\AA\ in the outer shells.


Nickel is the only element, out of the four considered, that posses ferromagnetic ordering at ambient conditions and has a rather high Curie temperature of 354$^{\circ}$C \cite{Legendre2011}. Therefore, the MTPs fitted to spin-dependent DFT calculations were used. Note that good agreement of non-magnetic calculations with the experimental EXAFS spectrum in Fig.\ \ref{fig:1}
is surprising and suggests some error cancellation---the numerical error of switching off magnetism is cancelled with the other error(s) whose source should be investigated.

\begin{figure}[t]
	\begin{center}
		\includegraphics[width=0.95\linewidth]{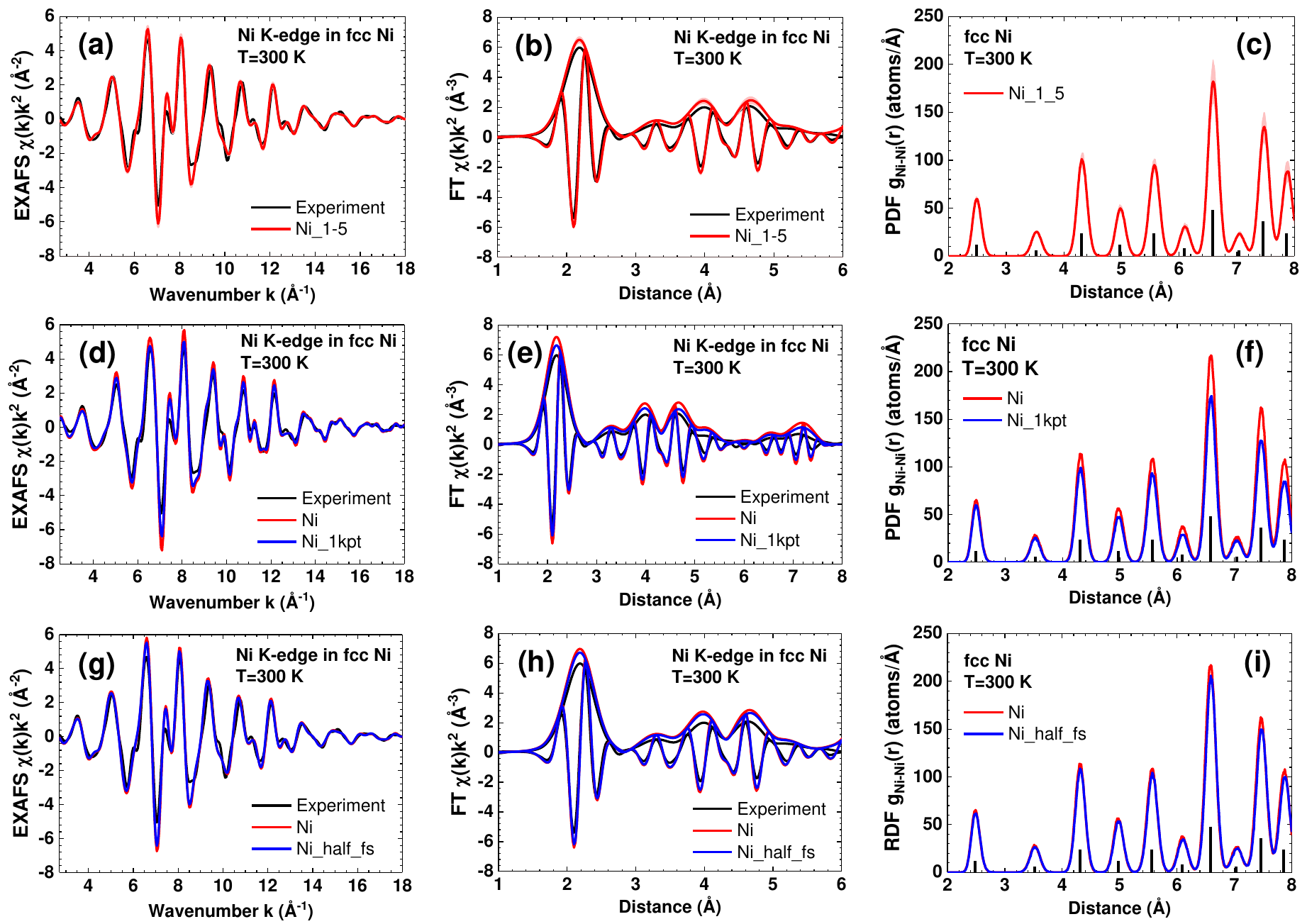}
	\end{center}
	\caption{Comparison of the experimental (black curves) and calculated Ni K-edge EXAFS $\chi(k)k^2$ spectra, their FTs and PDFs for fcc nickel.  The positions of crystallographic shells in fcc Ni are indicated by vertical bars in the PDF plots.   
	(a-c)  The calculated EXAFS spectrum is the average over five spectra generated for different MTPs. The scatter of calculated spectra is indicated.
	(d-f) Results for two MTPs.  We see that the spectrum calculated from spin-polarized MTP-predicted MD simulation (red curve) has the largest deviation compared to the result obtained with a single $\Gamma$-point (Ni\_1kpt, blue curve). 
	(g-i) The Ni K-edge EXAFS $\chi(k)k^2$ spectrum, their FTs and PDFs for fcc nickel generated with two different MD time steps of 1~fs (Ni, red curves) and 0.5~fs (Ni\_half\_fs, blues curves). Decreasing the MD time step affects little the results and hence does not explain the discrepancy between MTP and experiment.   
   }
	\label{fig:2}
\end{figure}

First, we evaluated the effect of the error between several MTPs trained to the same DFT model. To that end, we have fitted an ensemble of five different MTPs, and the average over five EXAFS spectra is compared with the experimental one in Fig.\ \ref{fig:2}a-c.
Although the five EXAFS spectra have a noticeable scatter, the experimental results are clearly outside the scope of this scatter.

Our next hypothesis was that the discrepancy was due to the numerical accuracy of DFT and MD, however, it did not stand as well---comparing our results with the less accurate $\Gamma$-point calculations (blue curves in Fig.\ \ref{fig:2}d-f) reveals that the more accurate results  (red curves in Fig.\ \ref{fig:2}d-f) are farther away from the experiment.
Also, decreasing the MD time step for 1~fs to 0.5~fs does not significantly change the results (blue curves in Fig.\ \ref{fig:2}g-i).

\begin{figure}[t]
	\begin{center}
		\includegraphics[width=0.95\linewidth]{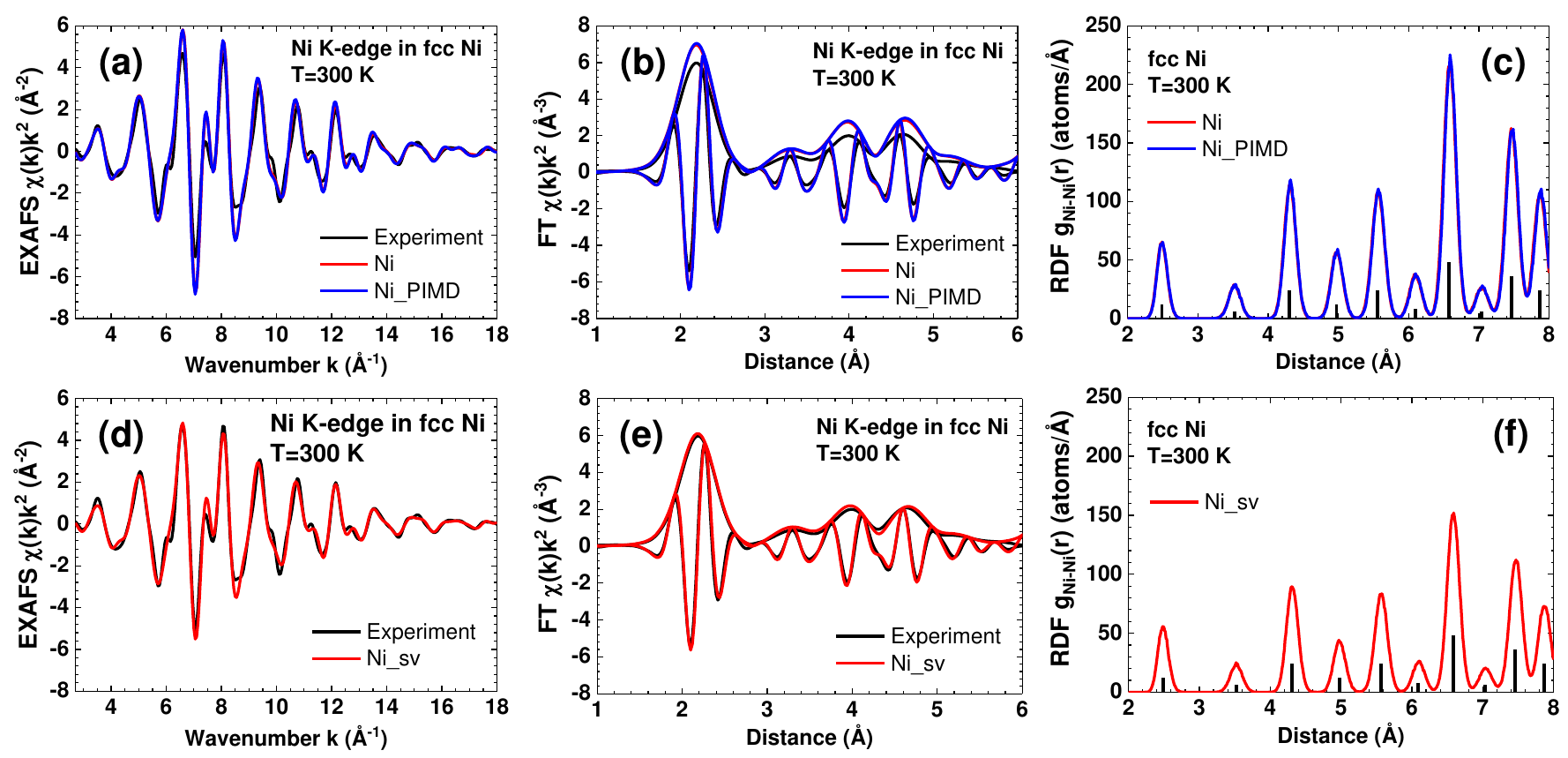}
	\end{center}
	\caption{Comparison of the experimental (black curves) and calculated Ni K-edge EXAFS $\chi(k)k^2$ spectra, their FTs and PDFs for fcc nickel. The positions of crystallographic shells in fcc Ni are indicated by vertical bars in the PDF plots.
		(a-c) Results for classical MD (red curves) and path-integral MD (PIMD) (blue curves) simulations. The nuclear quantum effects (zero-point energy) do not affect the results and hence do not explain the discrepancy between MTP and experiment.
	(d-f) Results calculated for fcc nickel with 18 electrons treated as valence in the PAW pseudopotentials are in nearly perfect agreement with the experiment.  }
	\label{fig:3}
\end{figure}

We then tested whether nuclear quantum effects (zero-point energy) can play a role in producing the discrepancy. Indeed, Ni is the lightest atom among the four, and the nuclear quantum effects should affect it the most. However, the results of path-integral molecular dynamics (PIMD) shown in Fig.\ \ref{fig:3}a-c prove that these effects are negligible at 300~K and cannot be the cause of the discrepancy.

Finally, we decided to choose an even more accurate (compared to the ``recommended'') pseudopotential for Ni, the one with 18 electrons, including the outer 3s orbital, treated explicitly as valence electrons (model Ni\_sv).
The simulations have been performed with the same DFT/MD parameters as for the other three elements (i.e., the same MD time step, $k$-point mesh, etc.).
The results are shown in Fig.\ \ref{fig:3}d-f.
One can see that the more accurate potential produces an almost perfect agreement with the experiment. This indicates that the accuracy of the pseudopotential was indeed the main source of discrepancy between the MTP and experimental results.

Atomic configurations obtained from the MD simulations can be used to calculate the 
mean-square relative displacements (MSRDs) $\sigma^2$ of atoms, known also as the Debye-Waller factors in EXAFS spectroscopy \cite{Beni1976,Dalba1997}. 
MSRDs determine the widths of the peaks in PDFs and  are responsible for the exponential damping of the EXAFS oscillations with increasing energy (wavenumber) and for their temperature dependence.  MSRD$_{ij}$ for the $i$-$j$ pair of atoms with the mean-square displacement (MSD) amplitudes MSD$_i$ and MSD$_j$ are related as ${\rm MSRD}_{ij}={\rm MSD}_i+{\rm MSD}_j-2 \phi \sqrt{{\rm MSD}_i} \sqrt{{\rm MSD}_j}$, where $\phi$ is a dimensionless correlation parameter \cite{Beni1976,Booth1995,Jeong2003}.  

For atoms located at large distances (in distant coordination shells), the correlation effects become negligible so that ${\rm MSRD} = 2 {\rm MSD}$ for monatomic metals, and, thus, a half of the MSRD provides an estimate of the mean-square displacement amplitude \cite{Jonane2016a,Jonane2018a}. Note that the MSD values are traditionally obtained from  diffraction measurements or lattice dynamics calculations, however, MD simulations and EXAFS data can be also used for this purpose \cite{Jonane2016a,Jonane2018a}.       

We found that PDFs for all four metals at $T=300$~K (Figs.\ \ref{fig:1} and \ref{fig:3})  can be well approximated by a set of Gaussian functions, whose widths give us an estimate for the MSRD values. The results for the first nine coordination shells of tungsten and molybdenum are reported in Table\ \ref{table1}. In both cases, 
the MSRD values for the nearest atoms include correlation effects due to metallic bonding,
however, at long distances, the MSRD increases and reaches saturation for $r$ above 5-6~\AA.   
Thus, for tungsten MSD(W)$\simeq 0.0020$~\AA$^2$\ and is in reasonable agreement compared to the experimental MSD(W)$=$0.0022~\AA$^2$\ in \cite{Paakkari1974} and the calculated MSD(W)$=$0.0023~\AA$^2$\ in \cite{Fine1939} and 0.0018~\AA$^2$\ in \cite{Dobrzynski1972}.
Our estimate for molybdenum is MSD(Mo)$\simeq 0.0025$~\AA$^2$\ compared 
to the experimental MSD(Mo)$=$0.0028~\AA$^2$\ in \cite{Paakkari1974} and the calculated 
 MSD(Mo)$=$0.0024~\AA$^2$\ in \cite{Dobrzynski1972}.
  
\begin{table}[t]
	\centering
	\caption{MSRD $\sigma^2$  for W--W and Mo--Mo interatomic distances $r$  in bcc W and Mo, respectively. $N$ is the coordination number. } 
	\footnotesize
	\begin{tabular}[t]{lcccc}
		\hline
		&  \multicolumn{2}{c}{W}  & \multicolumn{2}{c}{Mo}   \\		
		
		$N$ &	$r$ (\AA) & $\sigma^2$ (\AA$^2$) &	$r$ (\AA) & $\sigma^2$ (\AA$^2$) \\
		\hline
		8   &	2.74	& 0.0029   & 2.72  & 0.0041  \\
		6   &	3.16	& 0.0029   & 3.14  & 0.0038  \\
		12  &	4.47	& 0.0035   & 4.44  & 0.0048   \\
		24  &	5.24	& 0.0037   & 5.21  & 0.0048   \\
		8   &	5.47	& 0.0039   & 5.44  & 0.0051   \\
		6  & 	6.32	& 0.0037   & 6.28  & 0.0047   \\
		24  &	6.89	& 0.0038   & 6.85  & 0.0051   \\
		24 &	7.07	& 0.0041   & 7.02  & 0.0052   \\	
		24  &	7.74	& 0.0039   & 7.69  & 0.0053   \\
		\hline
	\end{tabular}\label{table1}
\end{table}%

Table\ \ref{table2} compares the values of the MSRD $\sigma^2$ for the first ten coordination shells of nickel in the Ni\_nonmag and Ni\_sv models.  One can see that the Ni\_sv model, which gives a better agreement with the experimental EXAFS data, has lower values of the MSRD factors in all coordination shells, which indicates somewhat more stiff interactions. Our estimate of MSD(Ni)$\simeq 0.0040$~\AA$^2$\ for the Ni\_sv model agrees reasonably well with the experimental MSD(Ni)$=$0.0049~\AA$^2$\ in \cite{Paakkari1974} and 0.00535~\AA$^2$\ in \cite{Jeong2003}.

The MSRD factors for copper are also shown in Table\ \ref{table2} for comparison.
The value of MSD(Cu)$\simeq 0.0062$~\AA$^2$\ is in good agreement with that 
measured by XRD MSD(Cu)$=$0.0065~\AA$^2$\  \cite{Wahlberg2016} and 
estimated from the effective field theory of phonons MSD(Cu)$=$0.0063~\AA$^2$\   \cite{Tomaschitz2021}.

\begin{table}[t]
	\centering
	\caption{MSRD $\sigma^2$  for Ni--Ni interatomic distances $r$  in the Ni\_nonmag and Ni\_sv models. $N$ is the coordination number. The MSRD values for fcc Cu are shown for comparison.  } 
	\footnotesize
	\begin{tabular}[t]{lccccc}
		\hline
		&	Distance & Ni\_nonmag  & Ni\_sv   &	Distance & Cu \\		
		
		$N$ &	$r$ (\AA) & $\sigma^2$ (\AA$^2$) & $\sigma^2$ (\AA$^2$) &	$r$ (\AA) & $\sigma^2$ (\AA$^2$)\\
		\hline
		12  &	2.49	& 0.0059   & 0.0055  & 2.55   & 0.0078  \\
		6  &	3.52	& 0.0090   & 0.0073  & 3.61   & 0.0115  \\
		24  &	4.31	& 0.0080   & 0.0070  & 4.42   & 0.0105   \\
		12  &	4.97	& 0.0083   & 0.0071  & 5.11   & 0.0110   \\
		24  &	5.56	& 0.0093   & 0.0077  & 5.71   & 0.0121   \\
		8  &	6.09	& 0.0085   & 0.0073  & 6.25   & 0.0108   \\
		48  &	6.58	& 0.0089   & 0.0077  & 6.75   & 0.0115   \\
		6  &	7.03	& 0.0103   & 0.0081  & 7.22   & 0.0133   \\	
		36  &	7.46	& 0.0097   & 0.0079  & 7.66   & 0.0117   \\
		24  &	7.86	& 0.0098   & 0.0081  & 8.07   & 0.0132    \\
		\hline
	\end{tabular}\label{table2}
\end{table}%


To conclude, we have tested the accuracy of the active learning algorithm of fitting the moment tensor potentials (MTPs) \cite{novikov2020-mlip,shapeev2016moment,gubaev2019-alloys} by comparing the EXAFS spectra computed from the molecular dynamics simulations with the experimentally obtained ones. To our knowledge, this is the first time a machine-learning potential for materials was validated using experimental EXAFS spectra. The employed approach is scalable to multicomponent alloys where EXAFS spectra can be independently measured for each element \cite{Zhang2018,FANTIN2020,Smekhova2021}.
	
The MTPs were fitted to DFT calculations with the recommended PAW pseudopotentials \cite{VASP1,VASP3,VASP4}. Out of the four tested elements, W, Mo, Cu, and Ni, the last one had the worst accuracy. We have then tested contributions of different physical and numerical sources of error, including the uncertainty of predictions of MTPs, the number of $k$-points for the DFT calculations, the size of the MD time step, the quantum nuclear effects, and magnetism. However, we found that the discrepancy between the computed and experimental EXAFS spectrum disappears only once we switch to the more accurate pseudopotential for Ni (from 10 to 18 valence electrons).
	
In a number of existing works, it was found that the difference between state-of-the-art machine-learning potentials and DFT is smaller than the difference between DFT and experimental predictions, which opens a route to multiscale simulations with ab initio accuracy.
Without running extremely expensive DFT calculations, we cannot make a similar conclusion from this study with certainty, however, our findings do support this statement. Hence, we suggest that the active learning of machine-learning potentials can be the right methodology to test and benchmark algorithms of DFT calculations.

\section*{Declaration of Competing Interest}
The authors declare that they have no known competing financial interests or personal relationships that could have appeared to influence the work reported in this paper.

\section*{Acknowledgements}

A.V.S. was supported by the Russian Science Foundation (grant number 18-13-00479).  Financial support provided by project No. 1.1.1.2/VIAA/l/16/147 (1.1.1.2/16/I/001) under the activity ``Post-doctoral research aid'' performed in the Institute of Solid State Physics, University of Latvia, is greatly acknowledged by D.B. Institute of Solid State Physics, University of Latvia as the Center of Excellence has received funding from the European Union's Horizon 2020 Framework Programme  H2020-WIDE-SPREAD-01-2016-2017-TeamingPhase2 under grant agreement No. 739508, project CAMART2.

\section*{Appendix. Supplementary materials}

Supplementary material associated with this article can be found, in the online version, at doi: . 



\end{document}